\documentstyle[12pt]{article}
\input{epsfig.sty}
\newcommand{\beq}{\begin{eqnarray}}
\newcommand{\eeq}{\end{eqnarray}}
\begin{document}
\title{$\Psi$ and $\Upsilon$ Production In pp Collisions at 7.0 TeV}
\author{Leonard S. Kisslinger\\
Department of Physics, Carnegie Mellon University, Pittsburgh PA 15213 USA\\
  kissling@andrew.cmu.edu\\
Debasish Das\\
Saha Institute of Nuclear Physics,1/AF, Bidhan Nagar, Kolkata 700064, INDIA\\
 dev.deba@gmail.com; debasish.das@saha.ac.in}
\date{}
\maketitle
\noindent
PACS Indices:12.38.Aw,13.60.Le,14.40.Lb,14.40Nd
\begin{abstract}
  This is an extension of recent studies for $\Upsilon(nS)$ and $\Psi(1S,2S)$ 
production at the LHC in pp collisions, $\sqrt{s}$=7.0 TeV, with the ALICE 
detector.
\end{abstract}

\noindent
Keywords:Heavy Quark Hybrid; LHC ALICE detector; rapidity cross-sections

\noindent
PACS Indices:12.38.Aw,13.60.Le,14.40.Lb,14.40Nd

\section{Differential rapidity cross sections for heavy quark state 
production at ALICE}

  This brief report is a continuation of our work on $\Upsilon(nS)$
production which was published recently\cite{kd13}. It is in anticipation
of the publication of new ALICE experimental results\cite{hxlalice13} on 
$J/\Psi(1S),\Psi(2S)$, and $\Upsilon(nS)$ production in the rapidity range 
$2.5 \leq y \leq 4.0$.

The differential rapidity cross section 
for $\lambda=0$ (dominant for $\Upsilon(nS),\Psi(nS)$ production) is given 
by~\cite{kmm11}
\beq
\label{dsig}
      \frac{d \sigma_{pp\rightarrow \Phi(\lambda=0)}}{dy} &=& 
     A_\Phi \frac{1}{x(y)} f_g(x(y),2m)f_g(a/x(y),2m) \frac{dx}{dy} \; ,
\eeq 
with $a= 4m^2/s$, $s=E^2$, $E=7.0$ TeV, $m=$ 5.0 GeV for Upsilon and 1.5 GeV 
for Charmonium states, $f_g$ the gluonic distribution function, and $x(y),
 \frac{dx}{dy}$ given in Refs\cite{kmm11},\cite{kd13}. For 
Upsilon, Charmonium $a=2.04 \times 10^{-6},1.8 \times 10^{-7}$. $\Phi$ in 
Eq(\ref{dsig}) is either $\Psi$ or $\Upsilon$, with  
$A_\Upsilon=1.74 \times 10^{-8}$ and $A_\Psi=6.46 \times 10^{-7}$.   

The gluonic distribution $f_g(x(y),2m)$ for the range of x needed for $E=7.0$
TeV is\cite{kmm11}
\beq
\label{fgupsilon}
         f_g(x(y))&=& 275.14 - 6167.6*x + 36871.3*x^2 \; .
\eeq

  The calculation of the production of $\Upsilon(3S)$ and $\Psi(2S)$ states
is done with the usual quark-antiquark model, and with the mixed heavy hybrid
theory\cite{lsk09}.
\clearpage
With these parameters we find for the differential rapidity cross sections
for $\Upsilon(1S)$ and $\Upsilon(2S)$ production as shown in the figures below.
\vspace{3cm}

\begin{figure}[ht]
\begin{center}
\epsfig{file=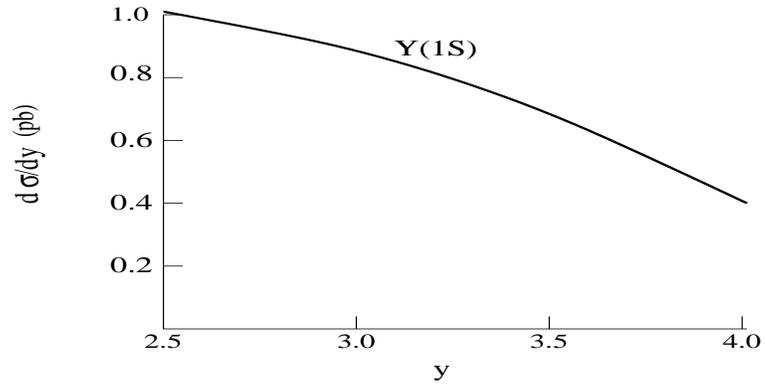,height=5cm,width=10cm}
\caption{d$\sigma$/dy for pp collisions at $\sqrt{s}$ = 7.0 TeV 
producing $\Upsilon(1S)$.}
\label{Figure 1}
\end{center}
\end{figure} 
\vspace{3cm}

\begin{figure}[ht]
\begin{center}
\epsfig{file=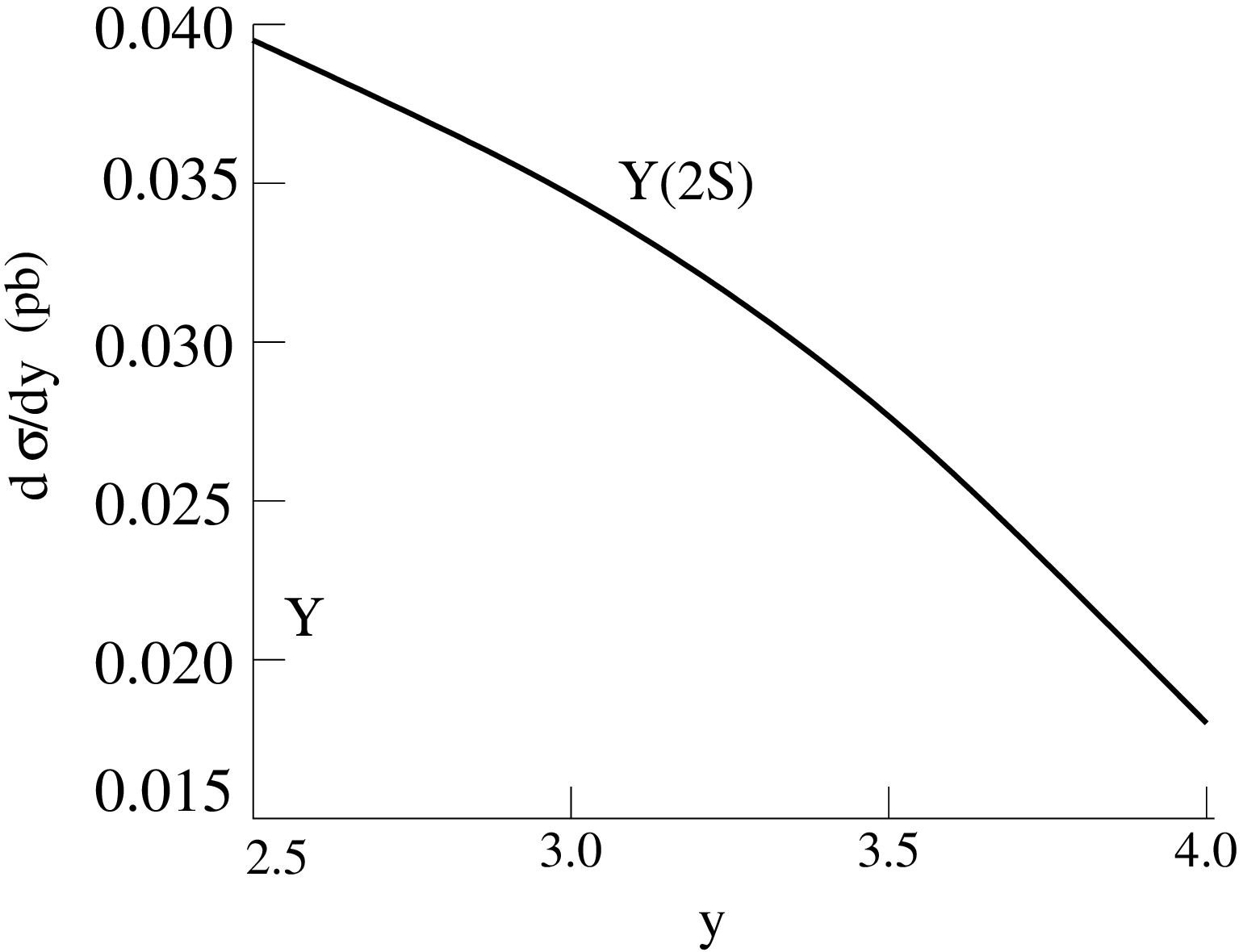,height=5cm,width=10cm}
\caption{d$\sigma$/dy for pp collisions at $\sqrt{s}$ = 7.0 TeV 
producing $\Upsilon(2S)$.}
\label{Figure 2}
\end{center}
\end{figure} 
\vspace{3cm}
\clearpage

The differential rapidity cross sections for $\Upsilon(3S)$ production 
with the hybrid and standard theories are shown in the figure below.
\vspace{3cm}

\begin{figure}[ht]
\begin{center}
\epsfig{file=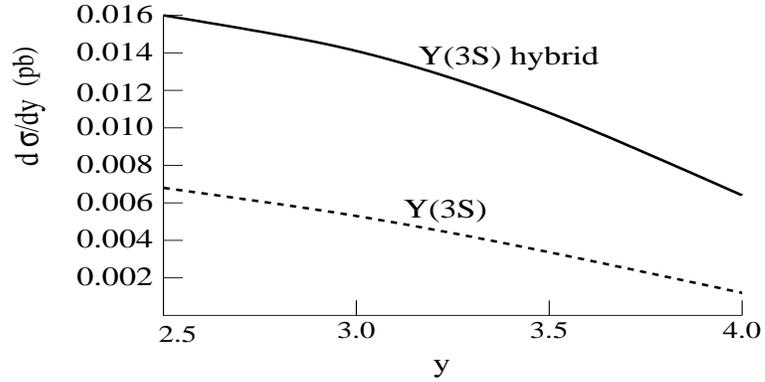,height=5cm,width=10cm}
\caption{d$\sigma$/dy for pp collisions at $\sqrt{s}$ = 7.0 TeV 
producing $\Upsilon(3S)$ for usual and hybrid theories.}
\label{Figure 2}
\end{center}
\end{figure} 
\vspace{2cm}

 The differential rapidity cross section for $J/\Psi(1S)$
is shown in the figures below.
\vspace{2cm}

\begin{figure}[ht]
\begin{center}
\epsfig{file=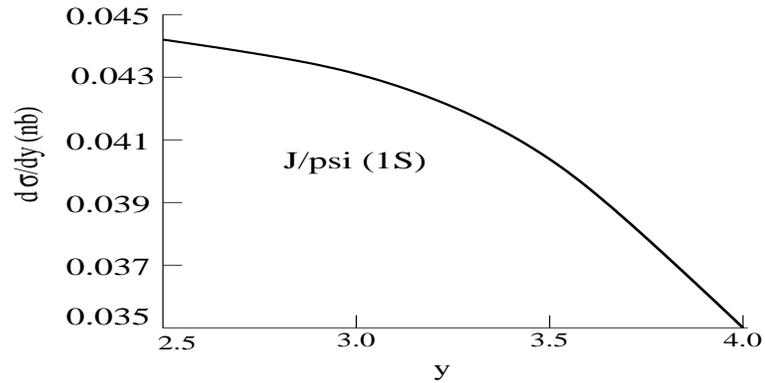,height=5cm,width=10cm}
\caption{d$\sigma$/dy for pp collisions at $\sqrt{s}$ = 7.0 TeV 
producing $J/\Psi (1S)$.}
\label{Figure 4}
\end{center}
\end{figure}
\clearpage
The differential rapidity cross sections for $\Psi (2S)$   production 
with the hybrid and standard theories are shown in the figure below.
\vspace{4cm}

\begin{figure}[ht]
\begin{center}
\epsfig{file=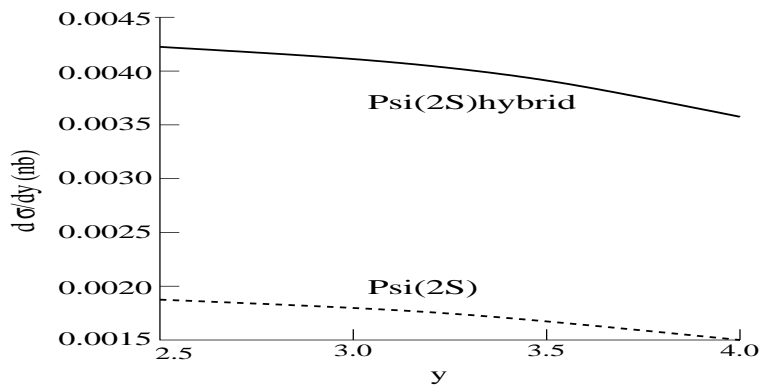,height=5cm,width=10cm}
\caption{d$\sigma$/dy for pp collisions at $\sqrt{s}$ = 7.0 TeV 
producing $\Psi (2S)$ for usual and hybrid theories.}
\label{Figure 5}
\end{center}
\end{figure} 

For $\Upsilon(3S)$ and $\Psi(2S)$ the standard $q\bar{q}$ prediction is
shown by dashed curves, while the prediction using the mixed hybrid 
theory\cite{lsk09} is shown with solid curves, with the difference
explained in Ref\cite{kmm11}.

\section{Conclusions}

  We expect that our results for the rapidity dependence of
d$\sigma$/dy shown in the figures can be useful for experimentalists
studing heavy quark production in p-p collisions at the LHC. It is also
a test of the validity of the mixed heavy quark hybrid theory, which we are 
using to test the creation of the Quark Gluon Plasma via relativistic
heavy ion collisions.
\newpage

\Large{{\bf Acknowledgements}}
\vspace{5mm}
\normalsize 

Author DD acknowledges the facilities of Saha Institute of Nuclear Physics, 
Kolkata, India. We thank Dr. Hugo Pereira Da Costa and his ALICE collaborators
for sending information about their forthcoming publication.

\end{document}